\magnification\magstep 1
\parskip 3pt plus 1pt minus 0.5pt
\def\a{\alpha}
\def\abs#1{{\left\vert#1\right\vert}}
\def\ai{\mathop{\rm Ai}}
\def\av#1{{\overline{#1}}}
\def\b{\beta}
\def\cm{_{\rm cm}}
\def\crr{\cr\noalign{\hrule}}
\def\d{{\rm d}}
\def\e{{\rm e}}
\def\eps{\varepsilon}
\def\euler{\gamma_{\scriptscriptstyle E}}
\def\frac#1#2{{#1\over #2}}
\def\frad#1#2{\displaystyle{#1\over #2}}
\def\i{{\rm i}}

\def\init{\tabskip 0pt}

\def\re{\mathop{\rm Re}}
\def\un{^{(1)}}
\def\z{^{(0)}}
{\parskip 0pt\parindent 0pt
\centerline{\bf ANOMALOUS DYNAMICAL SCALING AND BIFRACTALITY}
\medskip
\centerline{\bf IN THE ONE-DIMENSIONAL ANDERSON MODEL}
\bigskip\null\medskip
\centerline{by S. De Toro Arias$^{(1,2)}$\footnote{$^{\rm (a)}$}
{Email: sdetoro@ondine.unice.fr}
and J.M. Luck$^{(3)}$\footnote{$^{\rm (b)}$}
{Author to whom correspondence should be addressed.
E-mail: luck@spht.saclay.cea.fr}}
\bigskip\null\medskip
(1) Laboratoire de Physique de la Mati\`ere Condens\'ee\footnote{$^{(\star)}$}
{UMR 6622 of C.N.R.S.}, Universit\'e de Nice-Sophia-An\-ti\-po\-lis,
Parc Valrose, B.P. 71, 06108 Nice cedex 2, France.
\medskip
(2) C.E.A. Saclay, Service de Physique de l'\'Etat Condens\'e,
91191 Gif-sur-Yvette cedex, France.
\medskip
(3) C.E.A. Saclay, Service de Physique Th\'eorique,
91191 Gif-sur-Yvette cedex, France.
\vfill
{\bf Abstract.}
We investigate dynamical scaling properties of the one-dimensional
tight-binding Anderson model with a weak diagonal disorder,
by means of the spreading of a wave packet.
In the absence of disorder,
and more generally in the ballistic regime ($t\ll\xi_0$ in reduced units,
with $\xi_0$ being the localisation length near the band centre),
the wavefunction exhibits sharp fronts.
These ballistic fronts yield an anomalous time dependence
of the $q$-th moment of the local probability density,
or dynamical participation number of order $q$,
with a non-trivial exponent $\tau(q)$ for $q>2$.
This striking feature is interpreted as bifractality.
A heuristic treatment of the localised regime $(t\gg\xi_0)$
demonstrates a similar anomalous scaling, but with $\xi_0$ replacing time.
The moments of the position of the particle are not affected
by the fronts, and obey normal scaling.
The crossover behaviour of all these quantities
between the ballistic and the localised regime is described
by scaling functions of one single variable $x=t/\xi_0$.
These predictions are confirmed by accurate numerical data,
both in the normal and in the anomalous case.
\vfill
P.A.C.S.: 72.15.Rn, 71.23.An, 71.30.+h.\hfill S/98/041
\smallskip
To appear in Journal of Physics A\hfill T/98/053
\eject
}
\noindent{\bf 1 INTRODUCTION}
\smallskip

The Anderson localisation in a random potential
is now well understood,
at least as far as static or spectral properties are concerned~[1].
In the one-dimensional case, all eigenstates are exponentially localised,
with the localisation length $\xi(E)$ depending on energy $E$~[2,~3].
Dynamical aspects, concerning mostly the spreading of a wave packet,
have also recently attracted considerable interest~[4--6],
especially in connection with diffusion
in driven quantum systems and random band matrices~[7--9],
and more recently with the problem of two interacting particles~[10].

In this paper, we emphasise a novel and striking feature
of the dynamics of the one-dimensional tight-binding Anderson model.
The $q$-th moment of the local probability density,
or dynamical participation number of order $q$, to be defined in eq.~(1.4),
exhibits anomalous scaling and bifractal behaviour,
with a non-trivial exponent $\tau(q)$ for $q>2$.
This phenomenon seems to have been entirely overlooked so far.
It will be shown to take place in the absence of disorder,
and in the ballistic regime, in the localised regime,
and throughout the crossover between them.

To be more specific, we investigate the time-dependent wavefunction
of a tight-binding electron in one dimension, which obeys
$$
\i\frac{\d\psi_n(t)}{\d t}=\psi_{n+1}(t)+\psi_{n-1}(t)+v_n\psi_n(t),
\eqno(1.1)
$$
in reduced units,
such that lengths are measured in units of the lattice spacing,
and energies and inverse times in units of the hopping integral.
The diagonal site potentials $\{v_n\}$ are independent random variables,
drawn from a common distribution.
We choose the initial condition of a particle sitting at the site $n=0$
at the origin of times:
$$
\psi_n(0)=\delta_{n,0}.
\eqno(1.2)
$$
We characterise the spreading of the wave packet by the following quantities.

\noindent$\bullet$
Moments of the position of the particle:

$$
M_q(t)=\av{\langle\abs{n}^q\rangle}=\sum_n\abs{n}^q\av{P_n(t)}.
\eqno(1.3)
$$

\noindent$\bullet$
Moments of the probability density~(dynamical participation numbers):

$$
S_q(t)=\sum_n\av{\big(P_n(t)\big)^q}.
\eqno(1.4)
$$

\noindent
In the above definitions, the probability density reads
$$
P_n(t)=\abs{\psi_n(t)}^2,
\eqno(1.5)
$$
and the index $q$ is any real positive number, not necessarily an integer.
The bar denotes an average over the disorder,
i.e., over the distribution of the random site potentials $\{v_n\}$.
The conservation of the norm of the wavefunction ensures
that $M_0(t)=S_1(t)=1$ at all times $t\ge0$.
The most commonly considered quantities in the literature
are the mean squared position $M_2(t)$ and the participation number
(or inverse participation ratio) $S_2(t)$~[11--13].

Scaling properties are known to hold in the weak-disorder regime,
namely for a small enough random potential.
Assuming that the site potentials have zero average,
it is sufficient to characterise their distribution by its variance:
$$
\overline{v_n}=0,\qquad\overline{v_n^2}=\sigma^2\ll 1.
\eqno(1.6)
$$
To lowest order in perturbation theory,
the localisation length is maximal in the vicinity of the band centre,
where it scales as~[1--3]
$$
\xi_0\approx\frac{8}{\sigma^2}.
\eqno(1.7)
$$
This is the characteristic length scale
where the localisation phenomenon takes place.
The absence of an intermediate diffusive regime
between the ballistic one $(t\ll\xi_0)$ and the localised one $(t\gg\xi_0)$
is a peculiarity of the one-dimensional situation,
where there is no fundamental difference between the mean free path
and the localisation length~[14--18].
A more detailed description of the localisation length,
including its anomalous scaling near band edges,
will be recalled in section~3.1.

We shall be interested in the long-time behaviour
of the moments $M_q(t)$ and $S_q(t)$ defined above,
in the weak-disorder regime.
Scaling properties can be expected in this situation,
where both characteristic length scales $t$ and $\xi_0$
are simultaneously large.
The setup of this paper is as follows.
Section~2 contains a detailed analytical investigation of the problem
in the absence of disorder.
The wavefunction is given by a Bessel function,
which exhibits three different kinds of asymptotic behaviour
in the $(n,t)$ plane.
As a consequence, the moments $S_q(t)$ of the probability density
exhibit anomalous growth with a scaling exponent $\tau(q)$ for $q>2$.
This behaviour, interpreted as bifractality,
is expected to hold in the ballistic regime, namely for $t\ll\xi_0$,
where disorder will have hardly any effect on the ballistic motion
of the particle in the absence of disorder.
In section~3 we show on a heuristic basis that the same kind
of anomalous scaling, with the same exponent $\tau(q)$,
also takes place in the localised regime $(t\gg\xi_0)$,
but with the asymptotic width of the wave packet, of order $\xi_0$,
replacing time.
Finally, we demonstrate that all the quantities under consideration
exhibit scaling behaviour throughout the crossover
between the free (ballistic) and the localised (insulating) regimes,
involving universal scaling functions of the variable $x=t/\xi_0$.
This prediction is confirmed by accurate numerical data.
Section~4 contains a brief discussion.

\medskip
\noindent{\bf 2 ANALYTICAL RESULTS IN THE ABSENCE OF DISORDER}
\smallskip
\noindent{\bf 2.1 Description of the wavefunction}
\smallskip

In the absence of disorder,
the dynamics of the Anderson model can be investigated analytically.
Let us denote the various quantities with the superscript~(0)
in this limiting case.
The stationary tight-binding equation reads
$$
\psi\z_{n+1}+\psi\z_{n-1}=E\psi\z_n.
\eqno(2.1)
$$
Its eigenfunctions are the plane waves $\psi\z_n=\e^{\i np}$,
where momentum $p$ is related to energy $E$ by the dispersion relation
$$
E=2\cos p.
\eqno(2.2)
$$

We thus obtain an explicit expression for the time-dependent wavefunction,
$$
\psi\z_n(t)=\int_B\frac{\d p}{2\pi}\,\e^{\i np-2\i t\cos p}=\i^{-n}J_n(2t),
\eqno(2.3)
$$
where the momentum integral runs over the Brillouin zone $B=[-\pi,\pi]$.
The probability density at site $n$,
$$
P\z_n(t)=\abs{\psi\z_n(t)}^2=\big(J_n(2t)\big)^2,
\eqno(2.4)
$$
is thus the square of the Bessel function $J_n(2t)$,
whose argument is proportional to time,
while its order is the number of the site,
i.e., the distance travelled by the particle from its starting point.

It turns out that three regions have to be considered in the $(n,t)$ plane,
where the Bessel function admits different kinds of asymptotic behaviour.
The existence of these regions can be explained
by the following semi-classical argument.
The dispersion relation~(2.2) corresponds to the group velocity
$v=\d E/\d p=-2\sin p$.
As a consequence, for a wave packet initially peaked in momentum space
around some mean $p_*$, the centre of mass will move according to
the semi-classical law
$$
\langle n\rangle\approx-2t\sin p_*.
\eqno(2.5)
$$
We thus expect an allowed region $(\abs{n}<2t)$,
separated from a forbidden region $(\abs{n}>2t)$
by sharp fronts located at $n\approx\pm2t$.

The above heuristic picture can be made quantitative
by means of the following asymptotic formulas
in the theory of Bessel functions~[19].
As a matter of fact, the derivation of these formulas relies on the method
of steepest descent, with the saddle-point equation coinciding with
eq.~(2.5).

\noindent$\bullet$
Allowed region $(\abs{n}<2t)$:

For $n>0$, and with the notation
$$
n=2t\sin p\qquad(0<p<\pi/2),
\eqno(2.6)
$$
in agreement with the semi-classical law~(2.5), we have
$$
P\z_n(t)\approx\frac{\sin^2\big(n(p+\cot p-\pi/2)+\pi/4\big)}{\pi t\cos p}.
\eqno(2.7)
$$
The probability density is thus the product of $1/t$ by a rapidly oscillating
amplitude, as expected in an allowed region.

\vfill\eject
\noindent$\bullet$
Forbidden region $(\abs{n}>2t)$:

For $n>0$, and with the notation
$$
n=2t\cosh\theta\qquad(\theta>0),
\eqno(2.8)
$$
we have
$$
P\z_n(t)\approx\frac{\exp\big(-2n(\theta-\tanh\theta)\big)}{4\pi t\sinh\theta}.
\eqno(2.9)
$$
The probability density thus decays exponentially,
as expected in a forbidden region.

\noindent$\bullet$
Transition region $(\abs{n}\approx2t)$:

The transition region, corresponding to the ballistic fronts,
turns out to extend over a spatial range of order $t^{1/3}$.
For $n>0$, and with the notation
$$
n=2t+t^{1/3}z,
\eqno(2.10)
$$
the probability density is approximated as
$$
P\z_n(t)\approx t^{-2/3}\big(\ai(z)\big)^2,
\eqno(2.11)
$$
with $\ai(z)$ being the Airy function.
The asymptotic behaviour of this function, namely
$$
\ai(z)\approx\left\{\matrix{
\pi^{-1/2}\abs{z}^{-1/4}
\sin\left(\frac{2}{3}\abs{z}^{3/2}+\frac{\pi}{4}\right)\hfill&
(z\to-\infty),\hfill\cr
\frac{1}{2}\pi^{-1/2}z^{-1/4}\exp\left(-\frac{2}{3}z^{3/2}\right)\hfill&
(z\to+\infty),\hfill\cr
}\right.
\eqno(2.12)
$$
respectively matches eqs.~(2.7) and~(2.9).

Throughout the transition region,
the probability density is the product of $t^{-2/3}$
by an amplitude which oscillates toward the allowed region $(z\to-\infty)$,
and falls off exponentially toward the forbidden region $(z\to+\infty)$.

The existence of these three regions is illustrated in Figure 1,
showing the probability density $P\z_n(t)$ against site number $n$,
for a time $t=100$.

\medskip
\noindent{\bf 2.2 Moments of the position}
\smallskip

We now turn to the analysis of the moments $M\z_q(t)$
of the position of the particle, in the absence of disorder.
These quantities are quadratic forms in Bessel functions,
so that their analysis is rather easy,
at least for even integer values of the index: $q=2k$.
We have indeed
$$
\eqalign{
M\z_{2k}(t)
&=\sum_n n^{2k}\big(J_n(2t)\big)^2\cr
&=\int_B\frac{\d p}{2\pi}\,\int_B\frac{\d p'}{2\pi}\,\e^{2\i t(\cos p'-\cos p)}
\underbrace{\sum_n n^{2k}\e^{\i n(p-p')}}
_{2\pi\left(-\i\frac{\d}{\d p}\right)^{2k}\delta(p-p')}\cr
&=\int_B\frac{\d p}{2\pi}
\,\e^{-2\i t\cos p}\left(-\frac{\d^2}{\d p^2}\right)^k\e^{2\i t\cos p}.\cr
}
\eqno(2.13)
$$
The integrand of the last expression can be expanded as a trigonometric
polynomial.
As a result, the moments $M\z_{2k}(t)$ are even polynomials of $t$,
with positive integer coefficients.
We have $M\z_0(t)=1$, as expected, and
$$
M\z_2(t)=2t^2,\quad M\z_4(t)=6t^4+2t^2,\quad M\z_6(t)=20t^6+30t^4+2t^2,
\quad\hbox{etc.}
\eqno(2.14)
$$

The long-time behaviour of the moments is obtained
by letting all the derivatives act on the exponential
in the last expression of eq.~(2.13).
We thus obtain
$$
M\z_{2k}(t)\approx a_{2k}\,t^{2k}\qquad(t\gg1),
\eqno(2.15)
$$
with
$$
a_{2k}=\int_B\frac{\d p}{2\pi}\,(4\sin^2 p)^k=\frac{(2k)!}{(k!)^2}.
\eqno(2.16)
$$
This estimate can be shown to hold true for any real $q>0$, namely
$$
M\z_q(t)\approx a_q\,t^q\qquad(t\gg1),
\eqno(2.17)
$$
with
$$
a_q=\frac{2^q}{\pi^{1/2}}
\,\frac{\Gamma\left(\frac{q+1}{2}\right)}{\Gamma\left(\frac{q+2}{2}\right)},
\eqno(2.18)
$$
where $\Gamma(z)$ denotes Euler's gamma function.

\medskip
\noindent{\bf 2.3 Moments of the probability density~(participation numbers)}
\smallskip

The analysis of the long-time behaviour
of the moments $S\z_q(t)$ of the probability density is slightly more involved.
Indeed these quantities are highly non-linear functionals of the wavefunction.
As the index $q$ gets larger,
they are more and more sensitive to large values of the wavefunction,
whereas the moments $M\z_q(t)$ of the position are not.

The asymptotic expressions~(2.7), (2.9), (2.11) show that
the probability density $P\z_n(t)$ scales as $1/t$
in the allowed region (bulk of the wavefunction),
over an extent of order $t$ sites,
while it scales as $t^{-2/3}$ in the transition region
(fronts of the wavefunction),
over an extent of order $t^{1/3}$ sites,
and it is negligible in the forbidden region (tails of the wavefunction).
Hence the bulk has a normal contribution to the moment $S\z_q(t)$,
scaling as $t^{-(q-1)}$, while the anomalous contribution of the fronts
scales as $t^{-(2q-1)/3}$.
This analysis therefore predicts the power-law behaviour
$$
S\z_q(t)\approx b_q\,t^{-\tau(q)}\qquad(t\gg1),
\eqno(2.19)
$$
with $\tau(q)$ being the smaller of both exponents, namely
$$
\tau(q)=\left\{\matrix{
q-1\hfill&\hbox{for }q<2\hfill&\hbox{(normal)},\hfill\cr\cr
\frad{2q-1}{3}\hfill&\hbox{for }q>2\hfill&\hbox{(anomalous)}.\hfill\cr
}\right.
\eqno(2.20)
$$

This prediction is summarised in Table~1.
It can be made more quantitative by the following analysis,
yielding the value of the prefactor $b_q$ of the formula~(2.19) in either case.

\noindent$\bullet$
Normal regime $(q<2)$:

In the normal regime,
the moment $S\z_q(t)$ is dominated by the bulk of the wavefunction,
corresponding to the allowed region.
The prefactor of eq.~(2.19) can be estimated from the expression~(2.7),
assuming that the arguments of the sine functions are uniformly distributed,
and transforming the sum over $n$ into an integral over $p$.
We thus obtain after some algebra $S\z_q(t)\approx b_q\,t^{-(q-1)}$,
in agreement with eq.~(2.20), with
$$
b_q=\frac{2}{\pi^q}
\,\frac{\Gamma\left(\frac{2-q}{2}\right)}{\Gamma\left(\frac{3-q}{2}\right)}
\,\frac{\Gamma\left(q+\frac{1}{2}\right)}{\Gamma(q+1)}\qquad(q<2).
\eqno(2.21)
$$

\noindent$\bullet$
Anomalous regime $(q>2)$:

In the anomalous regime,
the moment $S\z_q(t)$ is dominated by the fronts of the wavefunction,
corresponding to the transition region.
The prefactor of this moment can then be estimated by using
the expression~(2.11), and transforming the sum over $n$
into an integral over $z$.
The outcome again agrees with eq.~(2.20), and yields
$$
b_q=2\int_{-\infty}^{+\infty}\d z\,\abs{\ai(z)}^{2q}\qquad(q>2).
\eqno(2.22)
$$
We have in particular
$$
b_3=0.073214.
\eqno(2.23)
$$

\noindent$\bullet$
Marginal case $(q=2)$:

This borderline case corresponds to the usual participation number $S_2(t)$.
The exponents of the contributions of the bulk and of the fronts
have the common value $\tau(2)=1$.
It is worth noticing that the prefactor $b_q$ diverges
as $b_q\approx3/(2\pi^2\abs{q-2})$ as $q\to 2$ from both sides.
This can be checked for $q<2$ directly from eq.~(2.21),
and for $q>2$ from the behaviour~(2.12) of the Airy function as $z\to-\infty$.

The moment $S\z_2(t)$ turns out to exhibit a logarithmic correction
to its leading $1/t$ behaviour,
which can be analysed by means of the Mellin transformation.
The function $S\z_2(t)$ and its Mellin transform $m(s)$ are related by
$$
m(s)=\int_0^\infty t^{s-1}\,S\z_2(t)\,\d t,\qquad
S\z_2(t)=\int\frac{\d s}{2\pi\i}\,t^{-s}m(s),
\eqno(2.24)
$$
for $\re s$ positive and small enough.
We have
$$
\eqalign{
S\z_2(t)
&=\sum_n\big(J_n(2t)\big)^4\cr
&=\int_B\frac{\d p_1}{2\pi}\dots\int_B\frac{\d p_4}{2\pi}
\,\e^{2\i t(\cos p_1+\cos p_2-\cos p_3-\cos p_4)}
\underbrace{\sum_n\e^{\i n(p_3+p_4-p_1-p_2)}}
_{2\pi\,\delta(p_1+p_2-p_3-p_4)}\cr
&=\int_B\frac{\d u}{2\pi}\int_B\frac{\d v}{2\pi}\int_B\frac{\d w}{2\pi}
\,\e^{-8\i t\sin u\sin v\sin w},
}
\eqno(2.25)
$$
where the last expression has been obtained by the change of variables
$$
\eqalign{
u&=(p_1-p_2+p_3-p_4)/4,\quad v=(p_1-p_2-p_3+p_4)/4,\cr
w&=(\pi-p_1-p_2)/2=(\pi-p_3-p_4)/2.
}
\eqno(2.26)
$$
The product structure of the last expression of eq.~(2.25)
makes it suitable to the closed-form evaluation of the Mellin transform $m(s)$.
We indeed obtain, for $0<\re s<1$,
$$
\eqalign{
m(s)&=\Gamma(s)\int_B\frac{\d u}{2\pi}\int_B\frac{\d v}{2\pi}
\int_B\frac{\d w}{2\pi}\,(8\i\sin u\sin v\sin w)^{-s}\cr
&=\Gamma(s)8^{-s}\cos\frac{s\pi}{2}
\left(\int_B\frac{\d u}{2\pi}\,\abs{\sin u}^{-s}\right)^3
=\Gamma(s)8^{-s}\cos\frac{s\pi}{2}
\left(\frac{\Gamma\left(\frac{1-s}{2}\right)}
{\pi^{1/2}\Gamma\left(\frac{2-s}{2}\right)}\right)^3.\cr
}
\eqno(2.27)
$$
The long-time behaviour of $S\z_2(t)$ is given by the double-pole singularity
of $m(s)$ at $s=1$, of the form
$$
m(s)=\frac{1}{2\pi^2}\left(\frac{1}{(1-s)^2}+\frac{6\ln 2+\euler}{1-s}
+\cdots\right),
\eqno(2.28)
$$
where $\euler$ denotes Euler's constant, hence
$$
S\z_2(t)\approx\frac{\ln t+6\ln 2+\euler}{2\pi^2t}\qquad(t\gg1).
\eqno(2.29)
$$
We have thus derived both the prefactor and the finite part
of the logarithmic correction of the dynamical participation number
$S\z_2(t)$ in the absence of disorder.
The finite part is a surprisingly large number:
writing the numerator of eq.~(2.29) as $\ln(t/t_0)$,
we have $1/t_0=64\exp(\euler)=113.989$.

\medskip
\noindent{\bf 2.4 Interpretation: bifractality of the probability density}
\smallskip

The scaling law~(2.19), (2.20) for the moments $S\z_q(t)$
of the probability density in the absence of disorder,
with its two branches of exponent $\tau(q)$,
can be interpreted within the multifractal formalism~[20].
Indeed, the wavefunction takes appreciable values over a number of lattice
sites of order $t$.
As a consequence, $1/t$ can be viewed as a short-distance cutoff
in the definition~(1.4) of the moments $S_q(t)$.

The scaling exponent $\tau(q)$ can be interpreted in terms of
generalised (R\'enyi) dimensions $D_q$ of the local probability density.
The relation $\tau(q)=(q-1)D_q$ yields
$$
D_q=\left\{\matrix{
1\hfill&\hbox{for }q<2\hfill&\hbox{(normal)},\hfill\cr\cr
\frad{2q-1}{3(q-1)}\hfill&\hbox{for }q>2\hfill&\hbox{(anomalous)}.\hfill\cr
}\right.
\eqno(2.30)
$$
The probability density can be alternatively characterised
by a multifractal spectrum $f(\alpha)$,
which is the Legendre transform of the exponent $\tau(q)$, according to
$$
\alpha=\frac{\d\tau}{\d q},\qquad f=q\frac{\d\tau}{\d q}-\tau.
\eqno(2.31)
$$
The expression~(2.20) yields the following results.
The normal branch $\tau(q)=q-1$, i.e., $D_q=1$,
for $q<2$ yields the point $(\alpha=1,f=1)$,
corresponding to the normal scaling of the bulk of the wavefunction,
while the anomalous branch $\tau(q)=(2q-1)/3$ for $q>2$ yields the point
$(\alpha=2/3,f=1/3)$,
corresponding to the anomalous scaling of the fronts of the wavefunction.

These results are summarised in Table~1.
We propose to call bifractality such a scaling behaviour,
with a normal and an anomalous component.

\medskip
\noindent{\bf 3 SCALING ANALYSIS IN THE GENERAL CASE}
\smallskip
\noindent{\bf 3.1 A reminder on band-edge anomalous scaling}
\smallskip

We first recall some results
on the scaling behaviour of the localisation length
in the presence of a weak diagonal disorder.
Inside the band of the pure system,
characterised by the dispersion relation~(2.2),
and in the weak-disorder regime $(\sigma^2\ll1)$,
the localisation length scales as~[1--3]
$$
\xi\approx\frac{8\sin^2 p}{\sigma^2}\approx\xi_0\,\sin^2 p.
\eqno(3.1)
$$

This leading-order perturbative prediction vanishes as $p\to 0$ or $p\to\pi$,
corresponding to the band edges, namely $E\to\pm2$.
This observation suggests that the localisation phenomenon
has something special near band edges.
This effect has been initially investigated by Derrida and Gardner~[21],
who indeed demonstrated the presence of anomalous scaling
in the localisation length $\xi(E)$ and the density of states $\rho(E)$.
These quantities behave near the upper band edge ($E\to2$, $\sigma\to0$) as
$$
\xi\approx\sigma^{-2/3}\Phi_1\big(\sigma^{-4/3}(E-2)\big),\qquad
\rho\approx\sigma^{-2/3}\Phi_2\big(\sigma^{-4/3}(E-2)\big),
\eqno(3.2)
$$
where the scaling functions $\Phi_1$ and $\Phi_2$ are known analytically.

Roughly speaking, the eigenstates whose energy lies near the band edges
have a localisation length of order $\sigma^{-2/3}$.
These states are therefore much more localised than typical eigenstates
within the band,
whose localisation length is of order $\xi_0\sim\sigma^{-2}$.
Only a small fraction of the whole spectrum,
of order $\sigma^{2/3}$ or $\xi_0^{-1/3}$,
consists of these anomalously localised eigenstates.

\medskip
\noindent{\bf 3.2 Heuristic analysis of the localised regime}
\smallskip

We now turn to a heuristic investigation of the moments of the position
and of the probability density, in the presence of a weak diagonal disorder,
and in the localised regime $(t\gg\xi_0\gg1)$.

Consider for definiteness
the Anderson model on a very long chain made of $N\gg1$ sites,
for a given realisation of the random potentials $\{v_n\}$.
Let $E^\a$ be the energy eigenvalues, labelled in some way by an integer $\a$,
and $\psi_n^\a$ be the corresponding eigenvectors.
We have
$$
\sum_n\psi_n^\a\psi_n^\b=\delta^{\a,\b},\qquad
\sum_\a\psi_m^\a\psi_n^\a=\delta_{m,n}.
\eqno(3.3)
$$
We define the centre-of-mass co-ordinate
$n\cm^\a$ and the localisation length $\xi^\a$ of every eigenstate as
$$
n\cm^\a=\sum_n n(\psi_n^\a)^2,\qquad
\xi^\a=\Big(\sum_n(n-n\cm^\a)^2(\psi_n^\a)^2\Big)^{1/2}.
\eqno(3.4)
$$

The initial condition~(1.2)
can be expanded as $\psi_n(0)=\delta_{n,0}=\sum_\a\psi_n^\a\psi_0^\a$.
As a consequence, the wavefunction reads at all times $t\ge0$
$$
\psi_n(t)=\sum_\a\e^{-\i E^\a t}\psi_n^\a\psi_0^\a.
\eqno(3.5)
$$

\medskip
\noindent{\it 3.2.1 Moments of the position}
\smallskip

Let us first take the example of the mean squared position,
for which eq.~(3.5) yields
$$
M_2(t)=\sum_{\a,\b}\e^{-\i(E^\a-E^\b)t}\psi_0^\a\psi_0^\b
\sum_n n^2\psi_n^\a\psi_n^\b.
\eqno(3.6)
$$
Our heuristic analysis of this expression will be based on
the following two hypotheses.

\noindent(A) Interference terms between different quantum states
can be neglected for large enough times.
Eq.~(3.6) thus becomes in the localised regime
$$
M_2(\infty)\approx\sum_\a(\psi_0^\a)^2\sum_n n^2(\psi_n^\a)^2.
\eqno(3.7)
$$

\noindent(B) Scaling properties of eigenstates can be modelled
by considering that the probability density $(\psi_n^\a)^2$
is roughly uniform over the range $\abs{n-n\cm^\a}<\xi^\a$.
This simple scaling hypothesis has been shown by analytical means
to hold in a variety of models, including random band matrices~[7,~8],
and the continuum Schr\"odinger equation in one dimension~[13].
It amounts to stating that single eigenstates
of the one-dimensional Anderson model do not exhibit multifractality,
in contrast with earlier claims based on numerical evidence~[22].

Consider first an eigenstate localised near the origin
$(\abs{n\cm^\a}\ll\xi^\a)$.
For such an eigenstate,
we have $\sum_n n^2(\psi_n^\a)^2\sim(\xi^\a)^2$,
while the prefactor $(\psi_0^\a)^2$ scales as $1/\xi^\a$.
Now, in a small energy interval $\Delta E$ around some energy $E$,
there are altogether $N\rho(E)\Delta E$ eigenstates,
among which only a finite number, of order $\xi(E)\rho(E)\Delta E$,
have $\abs{n\cm^\a}\sim\xi^\a$.
All these eigenstates bring comparable contributions,
of order $\big(\xi(E)\big)^2$, to the sum in eq.~(3.7).
It is worth noticing that the factor $\xi(E)$
in the number of relevant eigenstates
just compensates the prefactor $(\psi_0^\a)^2\sim1/\xi^\a$.
We are thus left with the estimate
$$
M_2(\infty)\sim\left\langle(\xi^\a)^2\right\rangle_\a,
\eqno(3.8)
$$
where the angular brackets denote an average
over the whole spectrum of eigenstates $\a$.
More explicitly,
$$
M_2(\infty)\sim\int\big(\xi(E)\big)^2\,\rho(E)\,\d E\sim\xi_0^2.
\eqno(3.9)
$$
We thus obtain the physically intuitive result that the mean squared position
saturates to a value of order $\xi_0^2$ for $t\gg\xi_0$.
Generalising the above argument, we get the asymptotic result
$$
M_q(\infty)\approx A_q\,\xi_0^q
\eqno(3.10)
$$
for all the moments of the position, deep in the localised regime.

This prediction can actually be made quantitative,
using results from the Russian literature~[14--17].
The long-time density correlation function has been calculated
in these references,
for the continuum Schr\"odinger equation with a weak white-noise potential.
If the initial wave packet is peaked in energy around some mean $E$,
one has $M_q(\infty)\approx p_q\big(2\xi(E)\big)^q$,
with the notation of ref.~[16],
where the amplitudes $p_q$ have been calculated analytically,
for integer values of $q$.
In the present situation, by averaging this prediction
over the whole spectrum of energies, we recover eq.~(3.10), with
$$
A_q=p_q\int_B\frac{\d p}{2\pi}\,(2\sin^2 p)^q.
\eqno(3.11)
$$
We have in particular
$$
A_2=\frac{3\zeta(3)}{4}=0.901543,\qquad
A_4=\frac{7\big(180\zeta(5)+\pi^4\big)}{128}=15.5343,\quad\hbox{etc.},
\eqno(3.12)
$$
where $\zeta$ denotes Riemann's zeta function.

\medskip
\noindent{\it 3.2.2 Moments of the probability density~(participation numbers)}
\smallskip

Let us now turn to the more interesting case of the moments $S_q(t)$
of the probability density, with $q=2, 3, \dots$ being an integer.
Eq.~(3.5) and hypothesis~(A) yield
$$
S_q(\infty)\approx\sum_{\a_1,\dots,\a_q}(\psi_0^{\a_1})^2\dots(\psi_0^{\a_q})^2
\sum_n(\psi_n^{\a_1})^2\dots(\psi_n^{\a_q})^2.
\eqno(3.13)
$$
Let us assume that the eigenstates $\a_1,\dots,\a_q$
are ordered according to increasing localisation lengths:
$\xi^{\a_1}<\xi^{\a_2}<\dots<\xi^{\a_q}$, and employ again hypothesis~(B).
If all the $q$ eigenstates have their centre-of-mass co-ordinates
close enough to the origin $(\abs{n\cm^{\a_k}}\ll\xi^{\a_k}$ for
$k=1,\dots,q)$, the product $(\psi_n^{\a_1})^2\dots(\psi_n^{\a_q})^2$
is nonzero where the $q$ eigenfunctions have a good common overlap.
This occurs in the intersection of all their ranges,
i.e., for $\abs{n}<\xi^{\a_1}$, where the product is of order
$1/\big(\xi^{\a_1}\xi^{\a_2}\dots\xi^{\a_q}\big)$,
while the sum over $n$ brings a factor of $\xi^{\a_1}$.
Here again, the number of relevant eigenstates in some energy range $\Delta E$
cancels out with the prefactors $(\psi_0^{\a_1})^2\dots(\psi_0^{\a_q})^2$,
whence the estimate
$$
S_q(\infty)\sim
\left\langle\frac{1}{\xi^{\a_2}\dots\xi^{\a_q}}\right\rangle_{\a_1,\dots,\a_q}.
\eqno(3.14)
$$

The averaging of this expression
over the eigenstates $\a_1,\dots,\a_q$ is more subtle than in the case of
eq.~(3.8), since negative powers of the localisation lengths are involved.
Hence the anomalously localised eigenstates near the band edges can,
and indeed will, play a role.

\noindent$\bullet$
If all the eigenstates $\a_1,\dots,\a_q$ belong to the bulk of the spectrum,
their localisation lengths scale as $\xi_0$,
and we obtain the normal prediction $S_q(\infty)\sim\xi_0^{-(q-1)}$.

\noindent$\bullet$
If, on the contrary, some of the eigenstates,
namely $\a_1,\dots,\a_m$, with $m\ge1$, belong to the band edges,
while $\a_{m+1},\dots,\a_q$ belong to the bulk of the spectrum,
the quantity to be averaged now scales as $\xi_0^{-(m-1)/3-(q-m)}$,
while the total fraction of such $q$-uples of eigenstates
is of order $\xi_0^{-m/3}$.
The optimal choice of the number $m$ of anomalous eigenstates is $m=q$,
hence the anomalous estimate $S_q(\infty)\sim\xi_0^{-(2q-1)/3}$.

The exponents of the above two estimates coincide
with those obtained in section 2.3 in the absence of disorder,
with the localisation length scale $\xi_0$ replacing time,
the bulk of the spectrum replacing the allowed region,
and the band edges replacing the ballistic fronts.
Generalising the above reasoning to non-integer values of the index $q$,
we predict the power-law behaviour
$$
S_q(\infty)\approx B_q\,\xi_0^{-\tau(q)}\qquad(\xi_0\gg1)
\eqno(3.15)
$$
for the participation numbers in the localised regime,
with the exponent $\tau(q)$ being given in eq.~(2.20) and in Table~1.
Finally, in analogy with the result~(2.29) in the absence of disorder,
a logarithmic correction of the form
$$
S_2(\infty)\approx\frac{\lambda\ln\xi_0+\mu}{\xi_0}\qquad(\xi_0\gg1)
\eqno(3.16)
$$
is expected in the marginal case $(q=2)$ of the usual participation number.
The values of the prefactors $B_q$, and $\lambda$ and $\mu$ cannot be predicted
by this heuristic analysis.

\medskip
\noindent{\bf 3.3 Scaling laws in the crossover regime}
\smallskip

So far, we have obtained two kinds of predictions concerning
the moments $M_q(t)$ of the position of the particle
and $S_q(t)$ of the probability density in the weak-disorder regime.
On the one hand,
the analytical results~(2.17), (2.19) obtained in the absence of disorder
are expected to hold more generally in the ballistic regime, i.e.,
for $1\ll t\ll\xi_0$.
On the other hand, a heuristic scaling analysis led us to
the predictions~(3.10), (3.15) in the localised regime, i.e.,
for $1\ll\xi_0\ll t$.

It turns out that the exponents involved in these results
always match between the ballistic and the localised regime,
both in the normal and in the anomalous case.
We are thus led to conjecture that the crossover
between these two limiting situations
is described by universal scaling functions of the single variable
$$
x=\frac{t}{\xi_0},
\eqno(3.17)
$$
throughout the scaling region where $t$ and $\xi_0$ are simultaneously large,
with the ballistic regime corresponding to $x\ll1$,
and the localised regime to $x\gg1$.
It is worthwhile to recall that the absence of an intermediate diffusive regime
between the ballistic one and the localised one
is a peculiarity of the one-dimensional geometry.

\medskip
\noindent{\it 3.3.1 Moments of the position}
\smallskip

We thus propose the following one-variable scaling law
for the moments of the position of the particle:
$$
M_q(t)\approx a_q\,t^q\,F_q(x),
\eqno(3.18)
$$
where the amplitudes $a_q$ are as in eq.~(2.18).
The result~(2.17) in the absence of disorder is recovered as $F_q(0)=1$,
while the estimate~(3.10) in the localised regime
implies the power-law fall-off
$$
F_q(x)\approx\frac{A_q}{a_q}\,x^{-q}\qquad(x\gg1).
\eqno(3.19)
$$

The scaling law~(3.18) can be shown to hold as $x\ll1$,
by means of a direct perturbative expansion in the random potentials~[23].
This approach, following the lines of refs.~[24,~3], yields
$$
F_q(x)=1-a\un_qx+\cdots\qquad(x\ll1),
\eqno(3.20)
$$
at least when the index $q=2k$ is an even integer, and in particular
$$
a\un_2=\frac{32}{3\pi}=3.39531,\qquad
a\un_4=\frac{1216}{135\pi}=2.86715,\quad\hbox{etc.}
\eqno(3.21)
$$

The analysis of the continuum Schr\"odinger equation~[17]
yields a behaviour similar to eq.~(3.20) at small $x$,
as well as a singular correction of relative order $(\ln x)/x$
to the leading power law~(3.19) at large $x$.
This logarithmic correction has been included in the analysis
of numerical data on large random band matrices~[9].

\medskip
\noindent{\it 3.3.2 Moments of the probability density~(participation numbers)}
\smallskip

Similarly, we postulate the following one-variable scaling laws
for the moments of the probability density
$$
S_q(t)\approx b_q\,t^{-\tau(q)}\,G_q(x)\qquad(q\ne2),
\eqno(3.22)
$$
where the amplitudes $b_q$ are as in eq.~(2.21) in the normal regime $(q<2)$,
and as in eq.~(2.22) in the anomalous regime $(q>2)$.
The result~(2.19) is recovered as $G_q(0)=1$,
while the estimate~(3.15) implies the power-law behaviour
$$
G_q(x)\approx\frac{B_q}{b_q}\,x^{\tau(q)}\qquad(x\gg1).
\eqno(3.23)
$$

In the marginal case of the usual participation number $S_2(t)$,
we expect a logarithmic correction of the form
$$
S_2(t)\approx\frac{\phi(x)\ln t+\chi(x)}{t}.
\eqno(3.24)
$$
The result~(2.29) in the absence of disorder is recovered as
$$
\phi(0)=\frac{1}{2\pi^2}=0.050660,\qquad
\chi(0)=\frac{6\ln 2+\euler}{2\pi^2}=0.239933,
\eqno(3.25)
$$
while the estimate~(3.16) in the localised regime implies
$$
\phi(x)\approx\lambda x,\qquad\chi(x)\approx(\mu-\lambda\ln x)x\qquad(x\gg1).
\eqno(3.26)
$$

\medskip
\noindent{\bf 3.4 Numerical results}
\smallskip

In order to confirm the scaling predictions of section 3.3,
we have performed direct numerical simulations of the dynamics
of the tight-binding Anderson model.
Introducing a finite time step $\eps$,
we have discretised eq.~(1.1) into the difference equation
$$
\psi_n(t+\eps)=\psi_n(t-\eps)-2\i\eps
\big(\psi_{n+1}(t)+\psi_{n-1}(t)+v_n\psi_n(t)\big).
\eqno(3.27)
$$
In the absence of disorder,
the dispersion relation of the corresponding stationary equation
between energy $E$ and momentum $p$ now reads
$$
\sin(\eps E)=2\eps\cos p.
\eqno(3.28)
$$
For a fixed momentum $p$, eq.~(3.28) has a first solution
$E^{(1)}=2\cos p+(4\eps^2/3)\cos^3p+\cdots$
that smoothly converges toward the continuous-time expression~(2.2),
and a second solution $E^{(2)}=\pi/\eps-E^{(1)}$ modulo $2\pi/\eps$,
corresponding to fast oscillations over a time scale $\eps$.

We have taken the initial condition~(1.2) at time $t=0$,
while we have chosen at time $t=\eps$
the Taylor expansion of the solution of eq.~(1.1),
to second order in $\eps$ included.
The only non-zero components in the full initial condition read
$$
\eqalign{
&\psi_0(0)=1,\qquad\psi_0(\eps)=1-\i v_0\eps-\big(1+v_0^2/2\big)\eps^2,\cr
&\psi_{\pm1}(\eps)=-\i\eps-\big(v_0+v_{\pm1}\big)\eps^2/2,\qquad
\psi_{\pm2}(\eps)=-\eps^2/2.
}
\eqno(3.29)
$$
This prescription reduces to the level of $\eps^3$
the amplitude of the fast oscillations related to the $E^{(2)}$ branch
of the dispersion relation~(3.28).

We want to emphasise that the discrete-time difference equation~(3.27)
has the very same physical contents as the continuous-time
differential equation~(1.1).
As an illustration of this point,
the mean squared position $M\z_2(t)$ still obeys
the ballistic law~(2.14), albeit with a prefactor given by
$$
a_2=\int_B\frac{\d p}{2\pi}\,\frac{4\sin^2p}{1-4\eps^2\cos^2p}
=\frac{1-(1-4\eps^2)^{1/2}}{\eps^2}=2+2\eps^2+\cdots
\eqno(3.30)
$$
The value $a_2=2$, characteristic of the continuous-time
equation~[see eqs.~(2.14), (2.16)],
is thus recovered, with a small correction in $\eps^2$.

We have performed numerical simulations of the difference equation~(3.27),
with a time step $\eps=0.05$,
so that the oscillations and other discretisation effects are negligible.
The random site potentials have been drawn from a uniform distribution
over the interval $-W/2<v_n<W/2$, so that $\sigma^2=W^2/12$ and
$$
\xi_0\approx\frac{96}{W^2}.
\eqno(3.31)
$$
The data to be presented below correspond to the following
ten values of the localisation length scale:
$$
\xi_0=25,\;35,\;50,\;70,\;100,\;140,\;200,\;280,\;400,\;560.
\eqno(3.32)
$$
For each value of $\xi_0$, the strength of disorder $W$
is taken from the perturbative relation~(3.31).
The measured quantities are averaged
over 1,000 independent realisations of the random potentials.
For each realisation, eq.~(3.27) is integrated up to
a time $t_{\rm max}=5\,\xi_0$, in order to enter into the localised regime,
and for a range of space $\abs{n}\le n_{\rm max}$,
with $n_{\rm max}=2.5\,t_{\rm max}=12.5\,\xi_0$,
in order to fully encompass the ballistic fronts.

\medskip
\noindent{\it 3.4.1 Moments of the position}
\smallskip

We have first checked the validity of the scaling law~(3.18)
for the moments $M_q(t)$ of the position of the particle,
as well as the analytical predictions
concerning the associated scaling functions,
on the examples of $q=2$ and $q=4$.
Figures~2 and 3 respectively
show log-log plots of numerical data for $M_2(t)$ and $M_4(t)$,
divided by their expressions~(2.14) in the absence of disorder,
against the scaling variable $x=t/\xi_0$.
The data, corresponding to the six largest values of eq.~(3.32) for $\xi_0$,
collapse in a nice way.
This demonstrates the existence of
the one-variable scaling functions $F_2(x)$ and $F_4(x)$.

The dashed lines show the linear correction~(3.20), (3.21)
of the scaling functions at small $x$,
as well as their asymptotic behaviour~(3.19) at large $x$,
namely $F_2(x)\approx0.4508/x^2$ and $F_4(x)\approx2.589/x^4$.
The full lines show the one-parameter phenomenological fits
$F_2(x)=(1+3.395\,x+0.154\,x\ln(x+1)+2.218\,x^2)^{-1}$
and $F_4(x)=(1+1.4336\,x+0.622\,x\ln(x+1)+0.6215\,x^2)^{-2}$.
These expressions incorporate the small-$x$ and large-$x$ behaviour
recalled just above.
The fitted parameters are the amplitudes
of the $x\ln(x+1)$ terms, reflecting the structure in $(\ln x)/x$
of the leading correction term at large $x$ discussed below eq.~(3.21).
The good quality of the fits shows the quantitative agreement
between analytical predictions and numerical data.
We have also checked that the amplitudes of the behaviour~(3.19), (3.20)
of the scaling functions at small and large $x$
are recovered within better than 10 percent
if they are left as free fitting parameters,
instead of being imposed as constraints.

\medskip
\noindent{\it 3.4.2 Moments of the probability density~(participation numbers)}
\smallskip

We now turn to the moments $S_q(t)$ of the probability density,
for which fewer analytical predictions are available.
We have first investigated the anomalous scaling laws of these moments
in the localised regime, on the example of $q=3$.
In order to check the power law~(3.15),
as well as the leading correction to it,
which can be expected to be of relative order $\xi_0^{-1/3}$,
we have plotted in Figure~4 the product $\xi_0^2\,S_3(\infty)$
against $\xi_0^{1/3}$.
The data points correspond to all the values of eq.~(3.32) for $\xi_0$.
For each $\xi_0$, the data for $S_3(t)$ in the range $1\le t/\xi_0\le5$
have been extrapolated, in order to get a reliable estimate for $S_3(\infty)$.
The error bars on the numbers obtained in this way are comparable
to the symbol size.
The plotted data nicely follow the least-square fit $y=2.46\,x-2.21$,
confirming thus both the power-law behaviour~(3.15)
and the anticipated nature of the correction term, and yielding the estimate
$$
B_3\approx2.5.
\eqno(3.33)
$$

In order to check the logarithmic behaviour~(3.16) of the
participation number in the localised regime,
we have plotted in Figure~5 the product $\xi_0\,S_2(\infty)$ against
$\ln\xi_0$.
The data points have been obtained as those of Figure~4.
They nicely follow the least-square fit $y=0.253\,x+0.642$,
confirming thus the behaviour~(3.16), and yielding the estimates
$$
\lambda\approx0.25,\qquad\mu\approx0.64.
\eqno(3.34)
$$

We have then determined the full one-variable scaling functions,
defined in eq.~(3.22), for the participation numbers
throughout the crossover from the ballistic to the localised regime.
We have again considered the example of $q=3$.
Figure~6 shows a log-log plot of $S_3(t)$,
divided by its behaviour~(2.19), (2.23) in the absence of disorder,
against the scaling variable $x=t/\xi_0$.
The data, corresponding to the six largest values of eq.~(3.32) for $\xi_0$,
again collapse in a nice way, demonstrating
the existence of the scaling function $G_3(x)$.
The asymptotic large-$x$ behaviour $G_3(x)\approx33.6\,x^{5/3}$,
shown as a dashed line, is accurately obeyed for values of the scaling variable
as small as $x\approx0.5$.
The full line shows the one-parameter phenomenological fit
$G_3(x)=(1-0.56\,x+67.8\,x^2)^{5/6}$,
with the fitted parameter being the amplitude of the middle term.

We end up with an investigation of the logarithmic behaviour~(3.24)
of the participation number $S_2(t)$,
for generic values of the scaling variable $x$.
To do so, for any fixed value of the ratio $x=t/\xi_0$,
we have performed a least-square fit of all the available data
for $tS_2(t)$ against $\ln t$,
the slope and the intercept of those fits respectively
yielding estimates for $\phi(x)$ and $\chi(x)$.
Figure~7 shows a log-log plot of the amplitude $\phi(x)$ thus obtained,
against the scaling variable $x$.
Dashed lines show the value of $\phi(0)$ given in eq.~(3.25),
and the asymptotic behaviour $\phi(x)\approx0.253\,x$~[see eq.~(3.26)].
The full line shows the one-parameter phenomenological fit
$\phi(x)=(0.00257+0.00282x\,+0.0639\,x^2)^{1/2}$,
with the fitted parameter being again the amplitude of the middle term.
Similar, albeit less accurate, numerical results
have been obtained for the function $\chi(x)$.

\medskip
\noindent{\bf 4 DISCUSSION}
\smallskip

The most salient outcomes of this work are summarised in Table~1.
In the absence of disorder,
and for an initially localised wave packet,
the probability density of a tight-binding particle
exhibits anomalous scaling and bifractality.
Indeed its moments of order $q$,
namely the dynamical participation numbers $S_q(t)$,
scale in time with a non-trivial exponent $\tau(q)$ for $q>2$.

To put it more boldly, a free quantum-mechanical particle is bifractal.
This striking feature is entirely due to the presence of ballistic fronts.
For the tight-binding model considered in this paper,
these fronts correspond to the transition region
in the theory of Bessel functions.
Both the existence of ballistic fronts and their width scaling as $t^{1/3}$
are actually general characteristics of difference equations,
with a bounded energy dispersion curve.
To the best of our knowledge, this bifractality phenomenon has been
overlooked so far.
Plots similar to our Figure~1,
showing the squared Bessel function with its ballistic fronts,
have been displayed and described e.g. in ref.~[25],
but without the authors noticing the relevance of the fronts.
The moments $M_q(t)$ of the position of the particle are not affected
at all by the presence of ballistic fronts.
Ref.~[5] contains a rigorous and general
discussion on the spreading of a wave packet.

For the Anderson model with a weak diagonal disorder,
the bifractal phenomenon persists throughout the different regimes
of the localisation phenomenon.
Indeed anomalous scaling of the participation numbers,
with the same exponent $\tau(q)$, holds in the ballistic regime $(t\ll\xi_0)$,
in the localised regime $(t\gg\xi_0)$,
and in the crossover between them $(t\sim\xi_0)$,
where these quantities obey scaling laws
involving the single variable $x=t/\xi_0$.
These scaling predictions have been confirmed quantitatively
by accurate numerical simulations.

The bifractal exponent $\tau(q)$ characterises
both the time decay of participation numbers in the ballistic case,
and their dependence on $\xi_0$ in the localised regime.
In the ballistic regime $(t\ll\xi_0)$,
this phenomenon is related to the existence of a finite upper band edge,
and to the uniform scaling of single eigenstates.
This absence of multifractality
is in turn related to the direct crossover between a ballistic
and a localised regime in the one-dimensional geometry,
without an intermediate diffusive phase.
These features have been established
by means of a variety of analytical techniques~[7,~8,~13--18].
In the localised regime $(t\gg\xi_0)$, bifractality is intimately related to
the Derrida-Gardner anomalous scaling of the
localisation length and of the density of states near band edges.

Two very different physical mechanisms
seem therefore to be responsible for the bifractal behaviour
observed in these two different regimes.
Let us underline that both phenomena have nevertheless
one important characteristic in common.
They take place near the band edges of the energy spectrum~(2.2)
in the absence of disorder,
where the two plane waves $\exp(\pm\i np)$ become identical.
This degeneracy of the plane-wave basis explains, at least qualitatively,
both the flatness of the dispersion curve around the upper band edge
(hence the existence of sharp ballistic fronts)
and the high sensitivity of eigenstates to a perturbation
(hence the Derrida-Gardner anomalous scaling).

The usual participation number $S_2(t)$ corresponds
to the borderline case $(q=2)$ between the normal and the anomalous regime.
This quantity exhibits a logarithmic correction,
either in time~[see eq.~(2.29)] or in $\xi_0$~[see eq.~(3.16)].
A similar logarithmic behaviour has been described~[26]
in the case of the mean return probability
$$
C(t)=\frac{1}{t}\int_0^t P_0(u)\,\d u.
\eqno(4.1)
$$
In the absence of disorder, this quantity has a logarithmic correction
to the naive scaling $C\z(t)\sim 1/t$,
which can be easily derived by the Mellin approach, yielding
$$
C\z(t)=\frac{1}{t}\int_0^t\big(J_0(2u)\big)^2\,\d u
\approx\frac{\ln t+4\ln 2+\euler}{2\pi t},
\eqno(4.2)
$$
a result very similar to eq.~(2.29).
This behaviour was shown in ref.~[26] to explain the occurrence
at several places in the literature of a fake non-trivial decay
exponent $\delta\approx 0.84$ for the mean return probability,
even in the absence of disorder,
as well as a variety of erroneous conclusions drawn from there.

Beyond the amusing phenomenon of bifractality
demonstrated in the one-dimensional Anderson model,
the present work underlines
that the moments of the position of a quantum-mechanical particle
and the moments of the associated probability density (participation numbers)
can exhibit different scaling laws, with unrelated dynamical exponents.
This outcome corroborates the recent general discussion~[5]
on the complexity of the scaling laws governing the dynamics
of quantum systems.

\medskip
\noindent{\bf Acknowledgements}
\smallskip

It is a pleasure for us to thank
J.L.~Pichard and X.~Waintal for interesting discussions,
and especially Y.~Fyodorov for an illuminating conversation.

\vfill\eject
{\parindent 0pt
\parskip 2pt plus 1pt minus 0.5pt
{\bf REFERENCES}
\medskip

[1] I.M. Lifshitz, S.A. Gredeskul, and L.A. Pastur, {\it Introduction to the
Theory of Disordered Systems} (Wiley, New-York, 1988).

[2] A. Crisanti, G. Paladin, and A. Vulpiani, {\it Products of Random Matrices
in Statistical Physics} (Springer, Berlin, 1992).

[3] J.M. Luck, {\it Syst\`emes d\'esordonn\'es unidimensionnels} (Collection
Al\'ea-Saclay, 1992).

[4] P. Sebbah, D. Sornette, and C. Vanneste, Phys. Rev. B {\bf 48} (1993),
12 506.

[5] R. Ketzmerick, K. Kruse, S. Kraut, and T. Geisel, Phys. Rev. Lett. {\bf 79}
(1997), 1959.

[6] B. Huckestein and R. Klesse, preprint cond-mat/9805038, and references
therein.

[7] A.D. Mirlin and Y.V. Fyodorov, J. Phys. A {\bf 26} (1993), L 551;
Y.V. Fyodorov and A.D. Mirlin, Phys. Rev. B {\bf 52} (1995), R 11 580.

[8] K. Frahm and A. M\"uller-Groeling, Europhys. Lett. {\bf 32} (1995), 385.

[9] F.M. Izrailev, T. Kottos, A. Politi, S. Ruffo, and G.P. Tsironis,
Europhys. Lett. {\bf 34} (1996), 441; F.M. Izrailev, T. Kottos, A. Politi,
and G.P. Tsironis, Phys. Rev. E {\bf 55} (1997), 4951, and references therein.

[10] F. Borgonovi and D.L. Shepelyansky, J. Phys. I France {\bf 6} (1996), 287;
S.N. Evangelou, S.J. Xiong, and E.N. Economou, Phys. Rev. B {\bf 54} (1996),
8469; K. Frahm, A. M\"uller-Groeling, and J.L. Pichard, Phys. Rev. Lett.
{\bf 76} (1996), 1509; S. De Toro Arias, J.L. Pichard, and X. Waintal
(in preparation).

[11] D.J. Thouless, Phys. Rep. {\bf 13} (1974), 93.

[12] F. Wegner, Z. Phys. B {\bf 36} (1980), 209.

[13] I.V. Kolokolov, Sov. Phys. JETP {\bf 76} (1993), 1099;
Europhys. Lett. {\bf 28} (1994), 193.

[14] V.L. Berezinskii, Sov. Phys. JETP {\bf 38} (1974), 620.

[15] A.A. Gogolin, V.I. Mel'nikov, and E.I. Rashba, Sov. Phys. JETP
{\bf 42} (1976), 168.

[16] A.A. Gogolin, Sov. Phys. JETP {\bf 44} (1976), 1003.

[17] E.P. Nakhmedov, V.N. Progodin, and Yu.A. Firsov, Sov. Phys. JETP
{\bf 65} (1987), 1202.

[18] P. Sheng (ed.), {\it Scattering and Locali\-zation of Classical Waves in
Random Media} (World Scientific, Singapore, 1990).

[19] A. Erd\'elyi (ed.), {\it Higher Transcendental Functions (The Bateman
Manuscript Project)} (McGraw-Hill, New-York, 1953).

[20] G. Paladin and A. Vulpiani, Phys. Rep. {\bf 156} (1987), 147; J. Feder,
{\it Fractals} (Plenum, New York, 1988).

[21] B. Derrida and E.J. Gardner, J. Phys. France {\bf 45} (1984), 1283.

[22] M. Schreiber and H. Grussbach, Phys. Rev. Lett. {\bf 67} (1991), 607;
S.N. Evangelou and D.E. Katsanos, J. Phys. A {\bf 26} (1993), L 1243.

[23] J.M. Luck (unpublished).

[24] B. Derrida and J.M. Luck, Phys. Rev. B {\bf 28} (1983), 7183.

[25] D.E. Katsanos, S.N. Evangelou, and S.J. Xiong, Phys. Rev. B {\bf 51}
(1995), 895.

[26] J.X. Zhong and R. Mosseri, J. Phys. Cond. Matt. {\bf 7} (1995), 8383.

\vfill\eject
{\bf CAPTIONS OF TABLE AND FIGURES}
\bigskip
{\bf Table~1:}
Summary of various characteristic features of the bifractality phenomenon.
\bigskip
{\bf Figure~1:}
Plot of the probability density $P\z_n(t)=\big(J_n(2t)\big)^2$
in the absence of disorder, against site number $n$, for a time $t=100$.
Arrows indicate the semi-classical ballistic fronts at $n=\pm200$.
\smallskip
{\bf Figure~2:}
Log-log plot of the ratio of mean squared position $M_2(t)$
to its value $M\z_2(t)$ in the absence of disorder,
against the scaling variable $x=t/\xi_0$.
Symbols: numerical data for various values of $\xi_0$.
Phenomenological fit (full line) and asymptotic behaviour (dashed lines)
are given in the text.
\smallskip
{\bf Figure~3:}
Same as Figure~2, for the fourth moment $M_4(t)$ of the position.
\smallskip
{\bf Figure~4:}
Plot of the product $\xi_0^2\,S_3(\infty)$ against $\xi_0^{1/3}$.
Symbols: numerical data for various values of $\xi_0$.
The least-square fit (full line) is given in the text.
\smallskip
{\bf Figure~5:}
Same as Figure~4, for the product $\xi_0\,S_2(\infty)$ against $\ln\xi_0$.
\smallskip
{\bf Figure~6:}
Log-log plot of the ratio of the participation number $S_3(t)$
to its value $S\z_3(t)$ in the absence of disorder,
against the scaling variable $x=t/\xi_0$.
Symbols and lines: as in Figure~2.
\smallskip
{\bf Figure~7:}
Log-log plot of the amplitude $\phi(x)$ of the participation number $S_2(t)$,
defined in eq.~(3.24), against the scaling variable $x=t/\xi_0$.
Symbols: numerical data.
Lines: as in Figure~2.
\vfill\eject
}
\centerline{\bf Table 1}
\medskip
$$
\vbox{\init\halign to 16truecm
{\strut#&\vrule#\tabskip=1em plus 2em&
\hfil$#$\hfil&\vrule#&
\hfil$#$\hfil&\vrule#&
\hfil$#$\hfil&\vrule#\tabskip 0pt\crr
&&\ &&\ &&\ &\cr
&&\hbox{component of bifractality}&&\hbox{normal}&&\hbox{anomalous}&\cr
&&\ &&\ &&\ &\crr
&&\ &&\ &&\ &\cr
&&\matrix{\hbox{relevant region of wavefunction}\cr
\hbox{in ballistic regime }(t\ll\xi_0)}&&
\matrix{\hbox{bulk}\cr\hbox{(allowed region)}\cr P_n(t)\sim t^{-1}}&&
\matrix{\hbox{fronts}\cr\hbox{(transition region)}\cr P_n(t)\sim t^{-2/3}}&\cr
&&\ &&\ &&\ &\crr
&&\ &&\ &&\ &\cr
&&\matrix{\hbox{relevant eigenstates}\cr
\hbox{in localised regime }(t\gg\xi_0)}&&
\matrix{\hbox{bulk of spectrum}\cr\xi\sim\xi_0\sim\sigma^{-2}}&&
\matrix{\hbox{band edges}\cr\xi\sim\sigma^{-2/3}}&\cr
&&\ &&\ &&\ &\crr
&&\ &&\ &&\ &\cr
&&\hbox{range of index }q&&q<2&&q>2&\cr
&&\ &&\ &&\ &\crr
&&\ &&\ &&\ &\cr
&&\hbox{exponent }\tau(q)&&q-1&&\frad{2q-1}{3}&\cr
&&\ &&\ &&\ &\crr
&&\ &&\ &&\ &\cr
&&\hbox{R\'enyi dimension }D_q&&1&&\frad{2q-1}{3(q-1)}&\cr
&&\ &&\ &&\ &\crr
&&\ &&\ &&\ &\cr
&&\hbox{multifractal spectrum}&&\alpha=1,\ f=1&&
\alpha=\frad{2}{3},\ f=\frad{1}{3}&\cr
&&\ &&\ &&\ &\crr
}}
$$
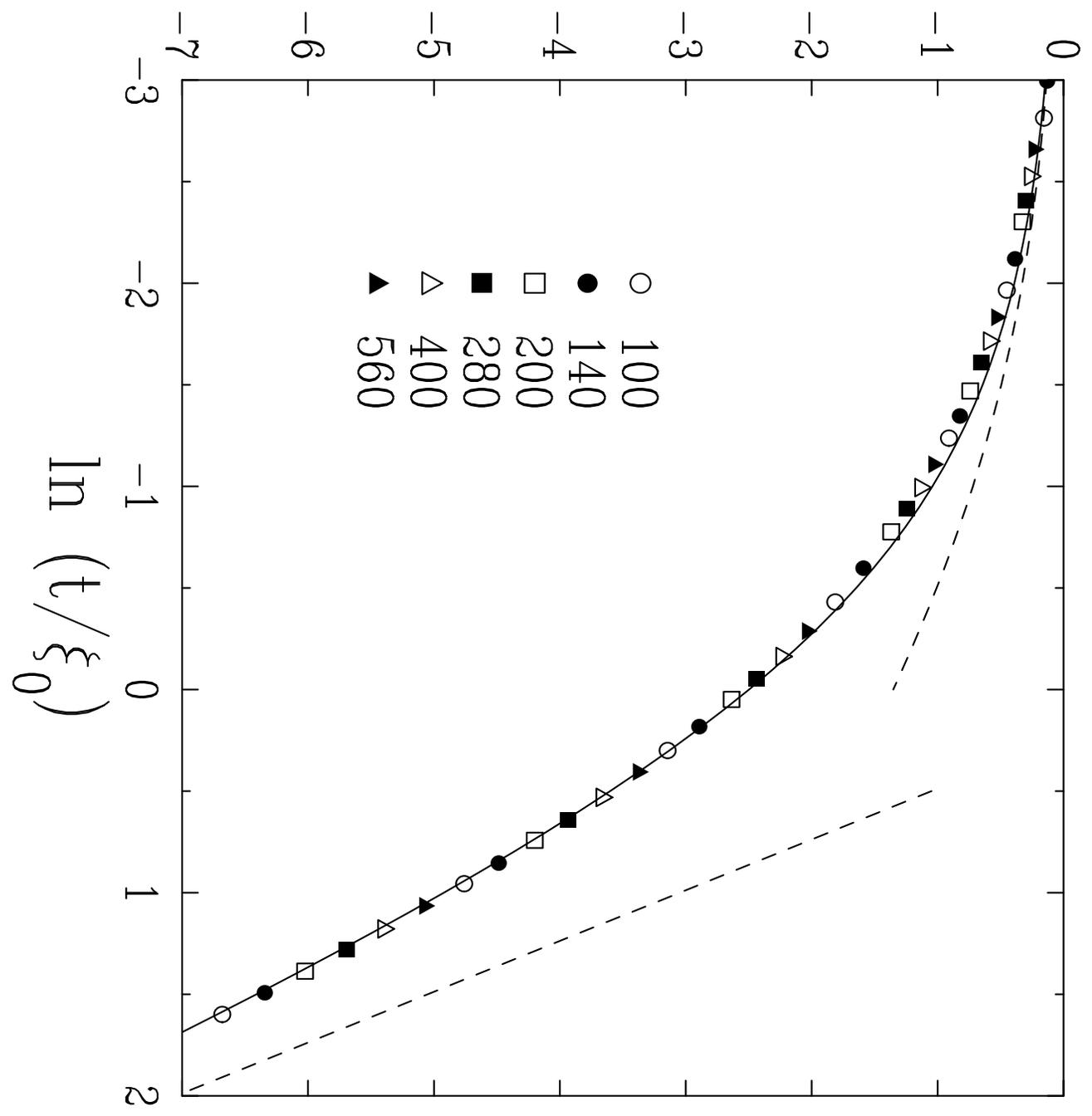
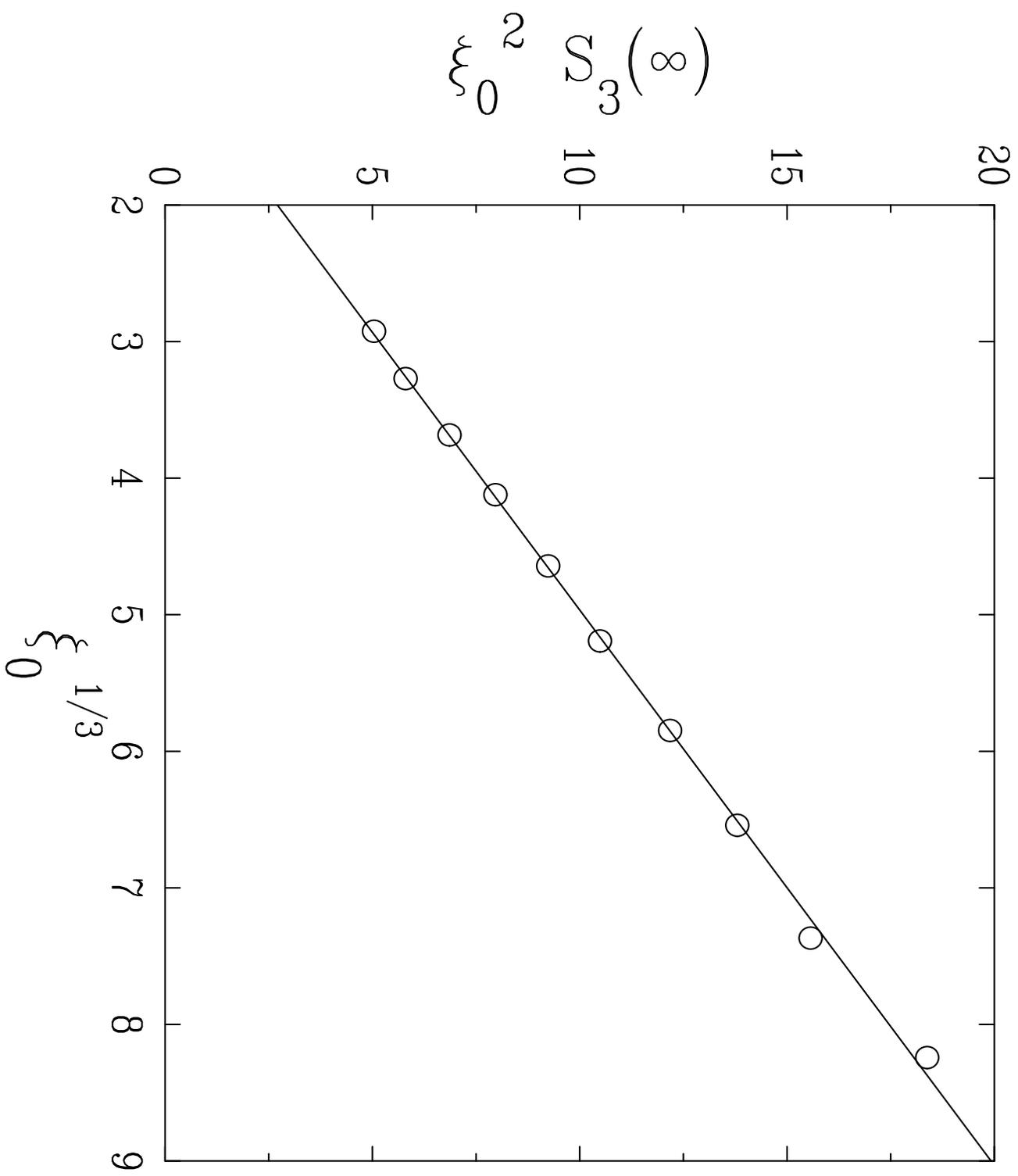
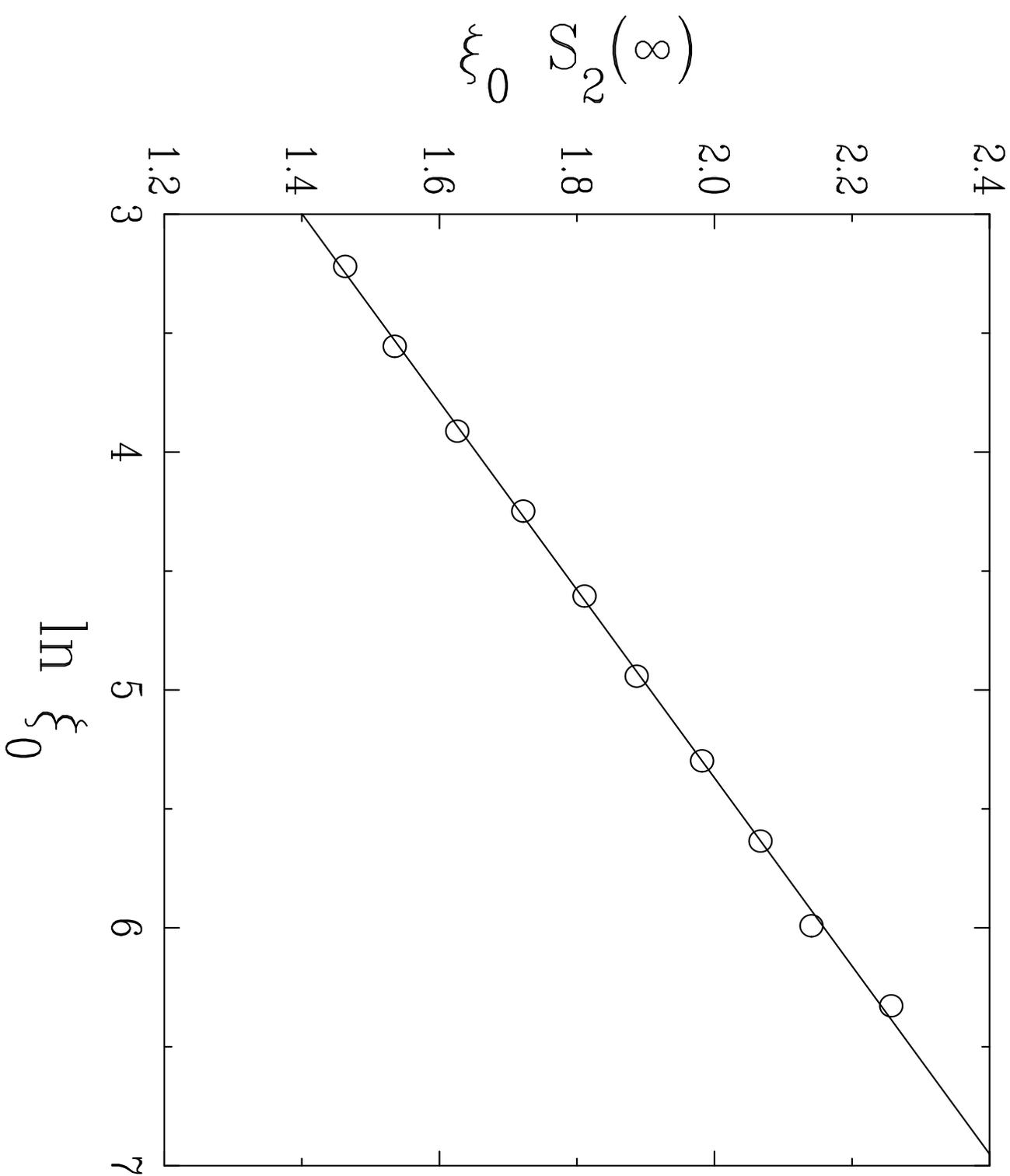
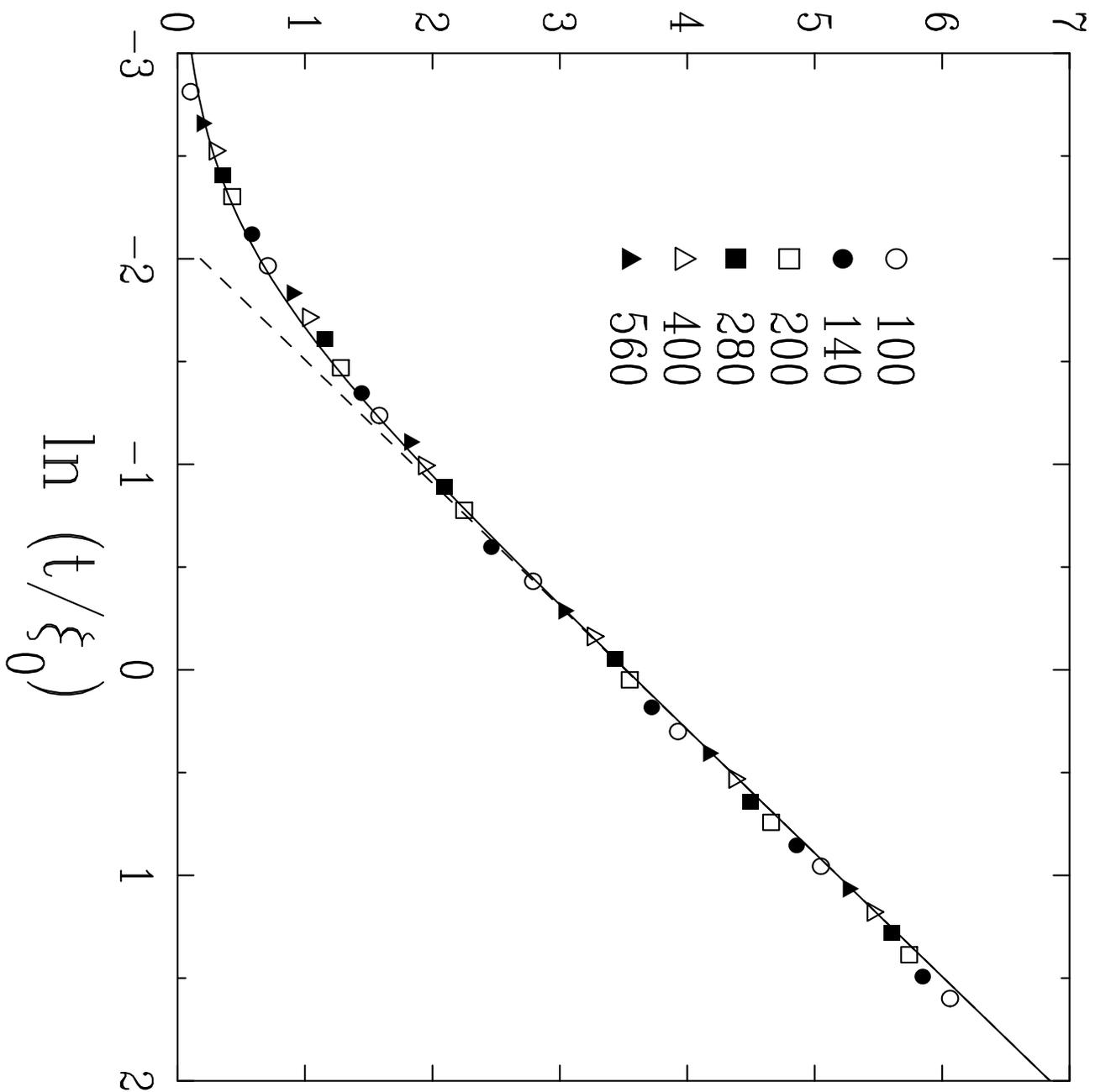
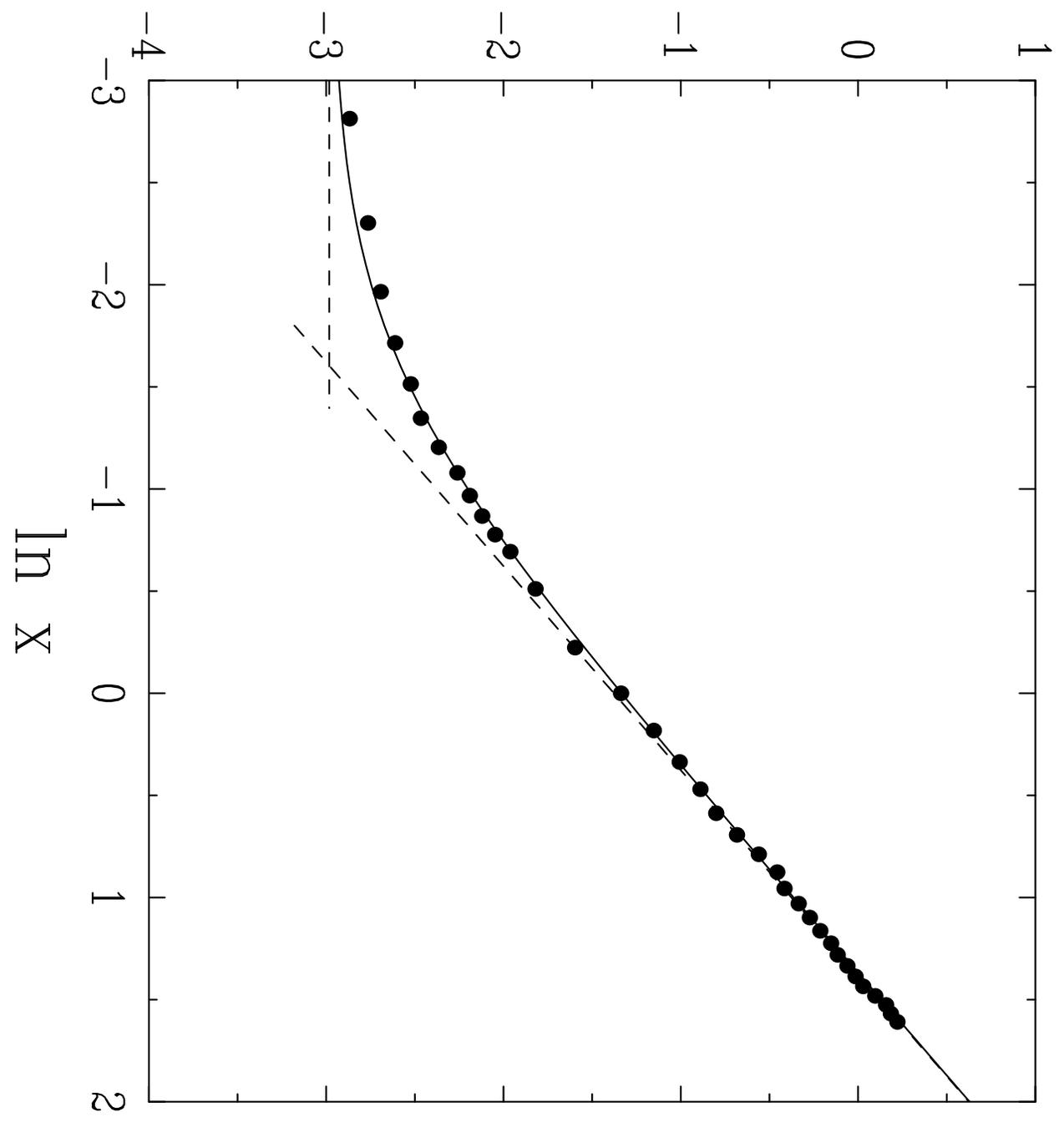
\bye